\documentclass[final]{article}

\PassOptionsToPackage{numbers,sort&compress}{natbib}

\usepackage[preprint]{neurips_2025}

\usepackage[utf8]{inputenc} 
\usepackage[T1]{fontenc}    
\usepackage{hyperref}       
\usepackage{url}            
\usepackage{booktabs}       
\usepackage{amsfonts}       
\usepackage{nicefrac}       
\usepackage{microtype}      
\usepackage{xcolor}         
\usepackage{color}
\usepackage{multirow}
\usepackage{tabularx}
\usepackage{graphicx}
\usepackage{natbib}
\usepackage{amsmath}
\usepackage{xspace}
\usepackage{colortbl}
\usepackage{enumitem}

\newcommand{\ours}{\textsc{Rexha}\xspace}

\graphicspath{{figs/}}

\title{Retrieval-Augmented Recommendation Explanation Generation with Hierarchical Aggregation}

\author{%
  Bangcheng Sun, Yazhe Chen, Jilin Yang, Xiaodong Li, Hui Li \\
  Key Laboratory of Multimedia Trusted Perception and Efficient Computing,\\
  Ministry of Education of China, Xiamen University, China \\ 
  \{36920231153231, 23020241154376, 22920212204262\}@stu.xmu.edu.cn \\
  \{xdli, hui\}@xmu.edu.cn \\
}

\begin{document}

\maketitle

\begin{abstract}
    Explainable Recommender System (ExRec) provides transparency to the recommendation process, increasing users' trust and boosting the operation of online services. With the rise of large language models (LLMs), whose extensive world knowledge and nuanced language understanding enable the generation of human-like, contextually grounded explanations, LLM-powered ExRec has gained great momentum. However, existing LLM-based ExRec models suffer from profile deviation and high retrieval overhead, hindering their deployment. To address these issues, we propose Retrieval-Augmented Recommendation Explanation Generation with Hierarchical Aggregation (\ours). Specifically, we design a hierarchical aggregation based profiling module that comprehensively considers user and item review information, hierarchically summarizing and constructing holistic profiles. Furthermore, we introduce an efficient retrieval module using two types of pseudo-document queries to retrieve relevant reviews to enhance the generation of recommendation explanations, effectively reducing retrieval latency and improving the recall of relevant reviews. Extensive experiments demonstrate that our method outperforms existing approaches by up to 12.6\% w.r.t. the explanation quality while achieving high retrieval efficiency. \label{abstract}

\end{abstract}

\section{Introduction}
\label{sec:intro}

The exponential growth of digital content and products across online platforms (e.g., Yahoo News, Amazon and TikTok) has intensified the information overload problem~\cite{BorchersHR98}, where users face overwhelming challenges in identifying relevant items amid vast information.
To mitigate cognitive burdens in decision-making, Recommender System (RecSys) has emerged as an essential tool and is prevalently incorporated into many online services.
RecSys effectively filters irrelevant information and delivers tailored recommendations.
This way, it not only helps users better find their desired targets but also facilitates the operation of platforms.
Hence, RecSys has attracted much attention and progressed actively over the past few decades~\cite{WuHWZW23,2022rsh}.

As users increasingly rely on recommendations for high-stakes decisions (e.g., financial investments and healthcare choices), mere predictive accuracy proves insufficient without explainable justification for recommendations 
This challenge has catalyzed the emergence of \underline{Ex}plainable \underline{Rec}ommender System (ExRec)~\cite{ZhangC20}.
ExRec explains the recommendation results to users and enhances the transparency of the decision-making process.
Therefore, it increases users' trust in RecSys, further boosting the operation of online services.

Prior works on ExRec mainly study how to generate interpretable explanations for user-item interactions~\cite{Ariza-CasabonaB24}.
For instance, some works adopt deep learning techniques like RNNs~\cite{LiWRBL17}, GNNs~\cite{WangCW22} and Transformer~\cite{LiZC20} for capturing intricate patterns in user behaviors and item attributes, enabling the generation of persuasive explanations.
Although these methods are effective in some cases, they naturally suffer from limited language competence as they are trained over the restricted recommendation data.
Hence, a branch of work on ExRec opts to use language models (e.g., BERT~\cite{PugoyK20}) pre-trained on vast, diverse textual corpora to generate human-like explanations.
This paradigm shift has gained momentum with the rise of large language models (LLMs), whose extensive world knowledge and nuanced language understanding enable the generation of human-like, contextually grounded explanations~\cite{LeiLYHL024,MaR024,LiZLCRKL25}.

XRec is one representative LLM-based ExRec~\cite{MaR024}. 
It enables LLMs to better understand complex user-item interaction patterns and offer explanations via integrating collaborative filtering (CF) signals. 
Building on this foundation, G-Refer~\cite{LiZLCRKL25} introduces a hybrid graph retrieval method to enhance ExRec by extracting both structural and semantic CF information from the user-item interaction graph. 
Despite these notable advancements, two critical limitations persist in LLM-powered ExRec: 
\textbf{(C1) Profile Deviation.} To assist with explanation generation, XRec and G-Refer construct user/item profiles by randomly sampling a small subset of user/item reviews, neglecting the broader contextual information embedded in the remaining data. This selective sampling introduces information bias, causing LLMs to generate explanations misaligned with holistic user preferences or item characteristics. 
\textbf{(C2) High Retrieval Overhead.} G-Refer employs the Dijkstra algorithm in path-level retrieval and can only be calculated on the CPU. It is not friendly to parallelism and has a high time complexity of $\mathcal{O}(N^2)$, resulting in a prohibitive retrieval cost.

\begin{figure}
    \centering
    \includegraphics[width=1\linewidth]{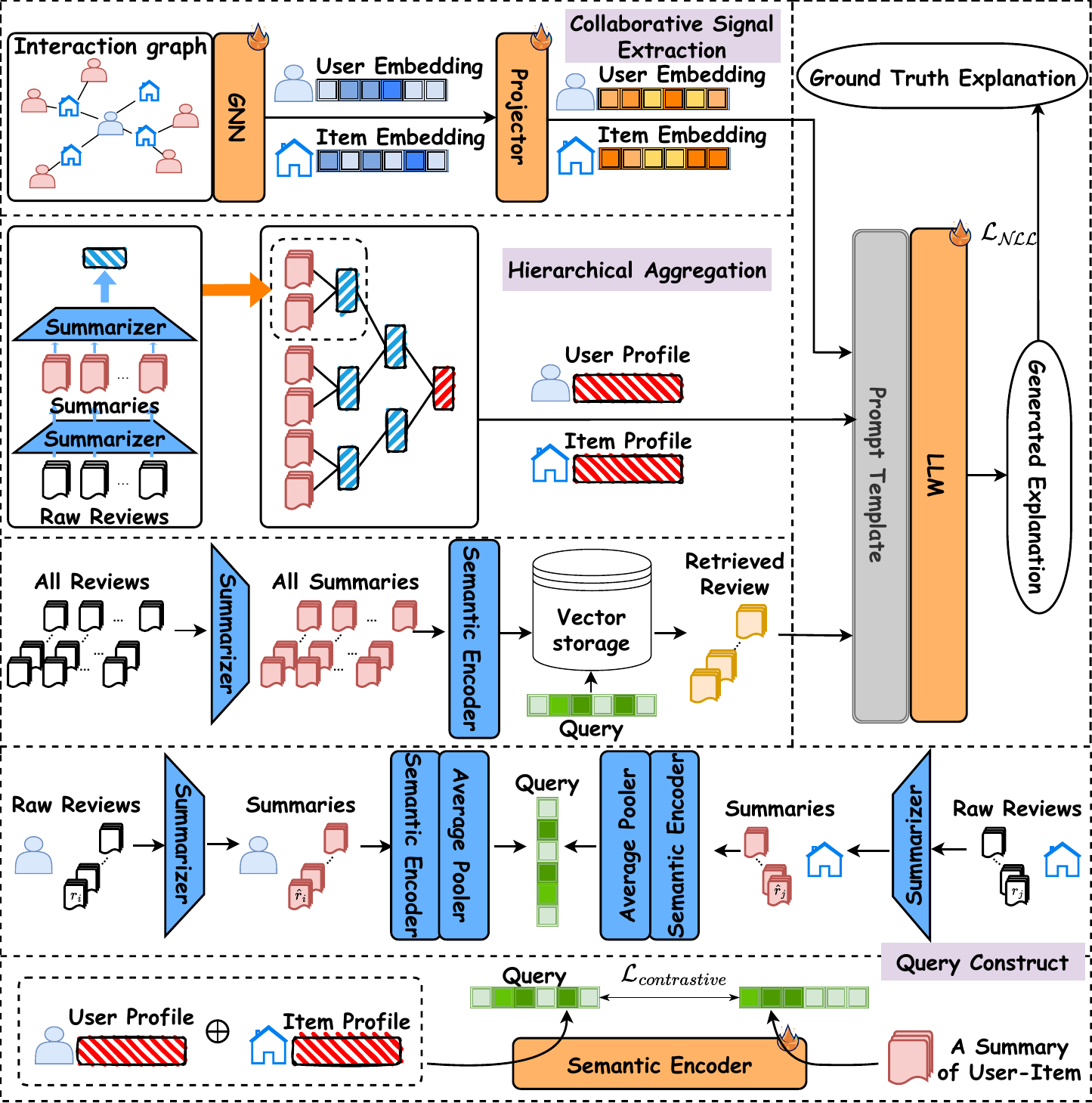}
    \caption{Overview of \ours. It contains three key components:(1) \textbf{Collaborative Signal Extraction} provides collaborative filtering information to LLMs. (2) \textbf{Hierarchical Aggregation} compresses and summarizes reviews layer by layer, finally construct textual profiles for user/item. (3) \textbf{Review Retrieval} module retrieves relative reviews to enhance LLM generating explanations.}
    \label{fig:overview}
\end{figure}

To alleviate the limitations of existing ExRec works, we present \underline{R}etrieval-Augmented Recommendation \underline{Ex}planation Generation with \underline{H}ierarchical \underline{A}ggregation (\ours), a framework designed to improve the credibility and efficiency of recommendation explanation generation. 
The contributions of this work are summarized as follows:
\begin{itemize}
    \item \textbf{Holistic User/Item Profiling:} Unlike prior methods (e.g., XRec and G-Refer) that construct incomplete profiles via random review sampling, we propose a hierarchical review aggregation module to systematically encode all user/item review data. This approach mitigates profile deviation (\textbf{C1}) by capturing nuanced user preferences and item characteristics through multi-layered summarization, ensuring LLMs generate explanations grounded in comprehensive contextual signals.
    
    \item \textbf{Efficient Retrieval via Pseudo-Document Queries:} To overcome the computational bottlenecks of  retrieval (\textbf{C2}), we introduce aggregated pseudo-document query generation. By synthesizing latent queries from user/item reviews, \ours efficiently retrieves semantically rich opinions from vectorized historical reviews, drastically reducing latency compared to prior LLM-powered ExRec.
    
    \item \textbf{Extensive and Rigorous Evaluation:} We have conducted extensive experiments on public datasets. The results demonstrate that \ours achieves significant performance gains, outperforming a range of state-of-the-art baselines by up to 12.6\% on recommendation explanation generation.
    
\end{itemize}

\section{Our Method \ours}
\label{sec:method}

Fig.~\ref{fig:overview} provides an overview of \ours.
The primary idea of \ours is to efficiently retrieve and harness reviews relevant to the target user-item interaction to enhance the credibility and personalization of the generated explanation.
To achieve this, \ours uses three main modules, collaborative signal extraction (Sec.~\ref{sec:coll}), hierarchical aggregation based profiling (Sec.~\ref{sec:ha}) and review retrieval for data augmentation (Sec.~\ref{sec:data_aug}), to prepare rich and beneficial background data for the target user-item interaction that is further fed into LLM for retrieval-augmented explanation generation (Sec.~\ref{sec:generation}).

\subsection{Collaborative Signal Extraction}
\label{sec:coll}

Collaborative information is crucial to model recommendation process~\cite{WuHWZW23}.
Constructing a user-item interaction graph and then capturing interaction patterns from the perspective of graph semantics and structure is a prevalent method, providing insights into understanding user preferences and item characteristics~\cite{WuSZXC23}.
Hence, in \ours, we leverage the capability of Graph Convolutional Networks (GCNs) to encode collaborative information in the user-item interaction graph into latent embeddings, complementing later LLM-based explanation generation by incorporating collaborative signals in recommendation scenarios. 

To be specific, we adopt LightGCN~\citep{DBLP:conf/sigir/0001DWLZ020-lightgcn} to encode the collaborative information:
\begin{equation}
    \mathbf{e}_{u}^{(l+1)}=\sum_{i\in\mathcal{N}_{u}}\frac{1}{\sqrt{|\mathcal{N}_{u}|}\sqrt{|\mathcal{N}_{i}|}}\mathbf{e}_{i}^{(l)},\quad\mathbf{e}_{i}^{(l+1)}=\sum_{u\in\mathcal{N}_{i}}\frac{1}{\sqrt{|\mathcal{N}_{i}|}\sqrt{|\mathcal{N}_{u}|}}\mathbf{e}_{u}^{(l)}
\end{equation}
where $\mathbf{e}^{(l)}_u$ and $\mathbf{e}^{(l)}_i$ denote the embedding of user $u$ and item $i$ after $l$-layer propagation, respectively. $\mathcal{N}_{u}$ denotes the set of users who have interacted with item $i$, and $\mathcal{N}_{i}$ indicates the set of items interacted by user $u$.

The user and item embeddings are extracted by averaging their embeddings from all GCN layers:
\begin{equation}
    \mathbf{e}_u=\sum_{l=0}^L\frac{1}{L+1}\mathbf{e}_u^{(l)},\quad\mathbf{e}_i=\sum_{l=0}^K\frac{1}{L+1}\mathbf{e}_i^{(l)}
\end{equation}
where $L$ denotes the number of layers. 
To align with the dimensionality of LLM's representation space, we further employ a three-layer feedforward neural network to project user and item embeddings. We utilize the negative log-likelihood (NLL) as our training loss, while the parameters of LLM are frozen.

\begin{equation}
    \mathcal{L} = -\frac{1}{N}\sum_{i = 1}^{N}\sum_{c = 1}^{C_i}y_{i,c}\log(\hat{y}_{i,c}).
\end{equation}

Here $N$ denotes the total number of user-item pairs to be explained, and $C$ denotes the token count in each explanations. $y_{i,c}$ and $\hat{y}_{i,c}$ correspond to the ground-truth and predicted tokens at position $c$ of $i$-th sample, respectively.

\subsection{Hierarchical Aggregation Based Profiling}
\label{sec:ha}
 
Item textual attributes (e.g., item title, item category, etc) contain important features for capturing item characteristics.
Nevertheless, as many items share similar textual attributes, there is insufficient information for constructing distinguishing profiles, making later explanation generation lack useful supporting inputs.
  
Instead, reviews written by humans reflect user preferences and item characteristics. 
Therefore, incorporating and summarizing fragmented human-written reviews using natural language as user and item profiles can enhance the richness and usefulness of the constructed profiles, providing better supporting inputs for LLM to generate persuasive and comprehensive recommendation explanations.

Existing LLM-based ExRec methods XRec and G-Refer utilize LLM as the summarizer.
Since processing all the reviews of a popular item may exceed the context length of LLM, they randomly sample a few reviews for constructing profiles, losing quite a lot of information beneficial for generating explanations. 
Furthermore, randomly sampled reviews lack connections to each other, making it more difficult for LLM to summarize and produce comprehensive and correct profiles. 
Despite that LLMs with a longer context window can be applied for profile summarization, the lost-in-middle issue~\cite{LiuLHPBPL24} still affects profile construction.

To conquer the above problem, we opt to construct user and item profiles via hierarchical aggregating their reviews instead of feeding reviews together to LLM.
\ours compresses and summarizes reviews layer by layer in parallel, and finally constructs a refined and informative profile.

\subsubsection{Raw Review Summarization}
\label{sec:summarize}

Given a user-item interaction, we conduct hierarchical aggregation twice: one for user profile and the other for item profile.
At the bottom of the hierarchical aggregation tree, each leaf contains one raw item review and each review is randomly placed in a leaf node.
When constructing a user profile, the raw review comes from one item interacted by that user.
When constructing an item profile, we use this item's reviews as raw reviews.
As shown in Fig.~\ref{fig:overview}, the first step is to aggregate reviews of two adjacent sibling nodes and use LLM to summarize the two raw reviews.

The primary motivation for employing LLM to summarize raw item reviews lies in addressing two critical challenges: excessive information redundancy and semantic ambiguity. 
By condensing user feedback into summaries, this approach establishes a robust semantic foundation that enhances the later hierarchical aggregation process.

Raw review data often suffers from verbosity, noise, and conflicting signals, such as redundant feature mentions (e.g., ``camera quality''), irrelevant details, or contradictory sentiments (``loves the design but hates the price''). 
Directly aggregating unprocessed text risks diluting actionable insights or introducing logical inconsistencies, which undermines both analytical precision and contextual coherence.
LLM excels at distilling these noisy inputs into concise, semantically dense representations. 
This process not only overcomes the limitation of context length but also extracts nuanced user preferences (e.g., ``prioritizes camera performance over battery life'') and item characteristics (e.g., ``high-resolution sensor with limited battery capacity''). 
The resulting summaries act as high-fidelity intermediaries, filtering out extraneous details while preserving critical semantic relationships.

Based on the above motivation, we design the instruction prompt to order the LLM to preliminarily summarize each raw reviews. The prompt can be found in Appendix~\ref{prompt_template}.
By combining the instruction prompt with a raw review $r_i$ as the input to LLM, we can leverage LLM to generate review summarization and similar processes are conducted in parallel for raw reviews.
This is similar to a $k$-ary-tree merging operation, where one raw review takes part in only one summarization.
Assume there are $m$ raw reviews at the bottom level. 
Then, after raw review summarization, there are $m/k$ brief but more information-dense summaries and they will be engaged in the next step of hierarchical aggregation.

\subsubsection{Hierarchical Aggregation}

When constructing an item profile for an item, after generating summaries from raw reviews, we concatenate every $p$ summaries into a group and conduct summarization similar to the summarization operation at the bottom level.
This process is conducted recursively until reaching the root level where we only have one summarization (i.e., the constructed item profile), forming a hierarchical aggregation tree as shown in Fig.~\ref{fig:overview}.
When integrated into the hierarchical aggregation process, the refined summaries from each level enable progressive fusion of granular features into the holistic profile. 
Similarly, a user profile can be constructed in a hierarchical manner.

\subsection{Review Retrieval for Data Augmentation}
\label{sec:data_aug}

\subsubsection{Review Retrieval Augmentation}

Hierarchical aggregation only utilizes reviews directly related to the target user/item to construct the user/item profile.
In practice, users often refer to the reviews of other similar items or consider reviews made by people with similar preferences before making decisions.
Such reviews may contain auxiliary and use information that may support their decisions.

Based on the above observation, \ours leverages an embedding model $f$ as the semantic encoder to transfer raw review summaries obtained in Sec.~\ref{sec:summarize} into representation vectors.
Then, for a target user-item interaction that requires explanation generation, \ours uses a retrieval query to find the most relevant raw review summaries based on cosine similarity. 
The query construction method will be introduced in the next section.
Finally, top-$q$ relevant raw review summaries to the target user-item interaction are listed as data augmentation for generating explanations.

\subsubsection{Retrieval Query Construction}
\label{query-generation}

Unlike general information retrieval scenarios, the recommendation system domain inherently lacks readily available natural language queries that can systematically retrieve review-based evidence for specific user-item interactions to enhance later explanation generation. 
In conventional recommendation frameworks, there exists no straightforward mechanism to translate implicit user preferences and item characteristics into search-compatible queries that directly enable contextual raw review summaries retrieval.

To address this problem, we propose two types of pseudo-document query construction strategies: the latent representation query and the profile query. The former leverages the global information of the user-item pair, aiming to retrieve diverse and informative reviews. The latter uses user and item profiles—constructed to emphasize high-frequency, long-term preference attributes—as queries. By retrieving based on these specialized profiles, the resulting reviews exhibit a more concentrated semantic distribution and are more closely aligned with the underlying preferences of the user or item.

\paragraph{Latent Representation Query} 
A latent representation query encodes the information of the target user-item interaction.
Concretely, \ours encodes all raw reviews of the target user and item using the embedding model $f$ (semantic encoder) and then aggregates all the encoded representations as the latent query for the target user-item interaction: 
\begin{align}
        &\hat{r}^u_i = f(r^u_i), \hat{r}^v_j = f(r^v_j), \\
        &\hat{R}^u = \{\hat{r}^u_1,\hat{r}^u_2,\dots,\hat{r}^u_n\},
        \hat{R}^v = \{\hat{r}^v_1,\hat{r}^v_2,\dots,\hat{r}^v_n\}, \\
        &\mathbf{Q}^u  = f(\hat{R}^u), \mathbf{Q}^v = f(\hat{R}^v), \\
        &\mathbf{q}^u  =\frac{1}{N}\sum_{q^u_i\in \mathbf{Q}^u}\mathbf{q}^u_i,
        \mathbf{q}^v  =\frac{1}{N}\sum_{q^v_i\in \mathbf{Q}^v}\mathbf{q}^v_i, \\
        &\mathbf{q}_{latent}  = \frac{\mathbf{q}^u + \mathbf{q}^v}{2}.
\end{align}
where $r^u_i$ is the raw review text written by the user, and $\hat{r}^u_i$ denotes its corresponding summarized opinion. Let $\hat{R}^u$ be the set of all user opinions $\hat{r}^u$. After mapping $\hat{R}^u$ to higher-dimension space by embedding model $f$, we obtain $\mathbf{Q}^u$, the set of embedding $\mathbf{q}^u$. $N$ denotes the number of opinions. The same applies to the item side.  

\paragraph{Profile Query} 
Since user and item profiles constructed in Sec.~\ref{sec:ha} contain precise and comprehensive information of user preferences and item characteristics, we design the profile query that directly uses the textual descriptions in profiles of the target user/item as the retrieval query.
Because user/item profiles misalign with raw review summaries due to filtering information during the hierarchical aggregation step. In this scenarios, fine-tuning the embedding model is necessary. 
We designed a contrastive learning approach that minimizes the distance between the user $u_i$ and item $v_j$ profiles $p_{u_i,v_j}$ and their corresponding opinions $\hat{r}_{u_i,v_j}$, while increasing the distance from irrelevant opinions.
\begin{equation}
    \begin{aligned}
        \mathcal{L}_{contrastive} = \frac{e^{sim(\mathbf{p}_{u_i,v_j},\hat{\mathbf{r}}_{u_i,v_j})/\tau}}
        {e^{sim(\mathbf{p}_{u_i,v_j},\hat{\mathbf{r}}_{u_i,v_j})/\tau}+\sum_{(i',j') \neq (i,j)}e^{sim(\mathbf{p}_{u_{i'},v_{j'}},\hat{\mathbf{r}}_{u_{i'},v_{j'}})/\tau}}.
    \end{aligned}
\end{equation}
where $sim(\cdot, \cdot)$ indicates cosine similarity, $\mathbf{p}_{u,v}$ and $\mathbf{r}_{u,v}$ denote embeddings of profiles and summarized reviews of the user-item pair, respectively.
The embedding $\mathbf{p}_{u,v}$ and $\mathbf{r}_{u,v}$ are computed as follows: 
\begin{equation}
    \mathbf{p}_{u,v} = f'(p_{u,v}),
    \mathbf{r}_{u,v} = f'(\hat{r}_{u,v}).
\end{equation}
where $f'$ denotes fine-tuned embedding model. $p_{u,v}$ and $\hat{r}_{u,v}$ denote the profile and summarized reviews of user-item pair $(u,v)$.

\subsection{Recommendation Explanation Generation}
\label{sec:generation}

For a target user-item interaction, we concatenate the corresponding constructed user and item profiles (Sec.~\ref{sec:ha}), the extracted user and item embeddings (Sec.~\ref{sec:coll}) and the relevant reviews for data augmentation (Sec.~\ref{sec:data_aug}) and use the prompt template shown in Fig.~\ref{fig:prompt_template} to instruct LLM to generate the recommendation explanation.

\section{Experiments}

\subsection{Experimental Settings}
\label{sec:experimental_settings}
\subsubsection{Datasets}
We conduct experiments on three public datasets from different domains: \textbf{Yelp}\footnote{\url{https://www.yelp.com/dataset/challenge}}, \textbf{Amazon-books}\footnote{\url{https://jmcauley.ucsd.edu/data/amazon/}}, and \textbf{Google-reviews}~\cite{DBLP:conf/acl/LiSM22,DBLP:conf/sigir/0003HLZM23}. 
More details of datasets can be found in Tab.~\ref{tab:datasets} in Appendix~\ref{datasets_detail}.

\subsubsection{Evaluation Metrics}

For evaluating the semantic explainability and stability of the generated explanation, we follow 
XRec~\cite{MaR024} and G-Refer~\cite{LiZLCRKL25} to employ a suite of metrics that assessing the semantic explainability and stability of generated explanations. 
Concretely, we use $\mathrm{GPT_{score}}$~\citep{DBLP:journals/corr/abs-2303-04048}, $\mathrm{BERT_{score}^{Precision}}$,$\mathrm{BERT_{score}^{Recall}}$ and  $\mathrm{BERT_{score}^{F1}}$~\citep{DBLP:conf/iclr/ZhangKWWA20} to measure the explainability. To evaluate the consistency, we also report the standard deviations of these metrics. The details of metric can be found in Appendix~\ref{appendix:metrics}.

\subsubsection{Baselines}
We compare our method with the following state-of-the-art methods, including NRT~\citep{LiWRBL17}, Att2Seq~\citep{ZhouLWDHX17}, PETER~\citep{DBLP:conf/acl/LiZC20-peter}, PEPLER~\citep{LiZC23a}, XRec~\citep{MaR024} and G-Refer~\citep{LiZLCRKL25}. The details of baseline can be found in Appendix~\ref{appendix:baselines}.

\subsubsection{Implementation Details}
\label{sec:implementation_details}

For the collaborative signal extractor module, we train LigntGCN with a learning rate of 1e-3 and a batch size of 1024. 
For hierarchical aggregation (HA) module, we set 4 reviews as a set to be summarized. 
For the retrieval module, we use llm-embedder model~\cite{DBLP:journals/corr/abs-2304-03728} to generate the embeddings of the reviews. 
For the generation module, we use the LLaMA-2-7B model as the base model. 
We set the learning rate, epochs, and batch size as 8e-4, 2 and 12, respectively. 
For inference, we set the temperature to 0 and the max output tokens as 128.
We run the experiments on a machine with 8 NVIDIA A800 GPUs.

\subsection{Overall Performance}
\label{sec:overall_performance}

\begin{table}[t]
    \centering
    \caption{Overall performance. Superscrips ``P'', ``R'' and ``F1'' respectively denote precision, recall, F1-score. Subscripts ``std'' denotes the standard deviation of each metric. Numbers in bold indicate the best results, while underlined numbers mean the second-best results. ``\ours-P'' and ``\ours-L'' correspond to \ours with latent representation queries and profile queries, respectively.}
    \label{overall-experiments}
    \resizebox{0.98\columnwidth}{!}{
    \begin{tabular}{l|cccc|cccc}
        \toprule
        \multicolumn{1}{c|}{\multirow{2}{*}{Models}} & \multicolumn{4}{c|}{\textbf{Explainability}$\uparrow$}& \multicolumn{4}{c}{\textbf{Stability}$\downarrow$} \\
        \cmidrule(lr){2-5} \cmidrule(lr){6-9}
        \multicolumn{1}{c|}{}                        & $\mathrm{GPT_{score}}$ & $\mathrm{BERT_{score}^P}$ & $\mathrm{BERT_{score}^{R}}$ & $\mathrm{BERT_{score}^{F1}}$ &  $\mathrm{GPT_{std}}$  & $\mathrm{BERT_{std}^{P}}$ & $\mathrm{BERT_{std}^{R}}$ & $\mathrm{BERT_{std}^{F1}}$  \\ 
        \midrule
        \multicolumn{9}{c}{\textbf{Amazon-books}}\\
        \midrule
        NRT           & 75.63         & 0.3444           & 0.3440             & 0.3443                       & 12.82        & 0.1804           & 0.1035           & 0.1321                          \\
        Att2Seq       & 76.08         & 0.3746           & 0.3624             & 0.3687                       & 12.56        & 0.1691           & 0.1051           & 0.1275                          \\
        PETER         & 77.65         & \underline{0.4279}           & 0.3799             & 0.4043            & 11.21        & 0.1334           & 0.1035           & 0.1098         \\
        PEPLER        & 78.77         & 0.3506           & 0.3569             & 0.3543                       & 11.38        & 0.1105           & 0.0935           & 0.0893                         \\
        XRec          & 82.57         & 0.4193           & 0.4038           & 0.4122                       & 9.60         & \textbf{0.0836}            & 0.0920          & \underline{0.0800}\\
        G-Refer (7B)  & \underline{82.70}         & 0.4076           & \textbf{0.4476}             & \underline{0.4282}                       & \underline{9.04}         & 0.0937              & \textbf{0.0845}           & 0.0820\\
        \rowcolor{gray!20}
        \ours-P       & \textbf{83.28}           & 0.3995 	    & 0.3752 	&0.3881 	& 9.69           & \underline{0.0844} &	\underline{0.0896} &	\textbf{0.0788}\\
        \rowcolor{gray!20}
        \ours-L         & 81.44     &\textbf{0.4722}        & \underline{0.4070}  &\textbf{0.4400}      &\textbf{8.95}&0.0908&0.0989&0.0862 \\

        \midrule
        \multicolumn{9}{c}{\textbf{Yelp}}\\
        \midrule
        NRT         &61.94          &0.0795             &0.2225             &0.1495                             &16.81              &0.2293             &0.1134     &0.1581          \\
        Att2Seq     &63.91          &0.2099             &0.2658             &0.2379                             &15.62              &0.1583             &0.1074     &0.1147         \\
        PETER       &67.00          &0.2102             &0.2983             &0.2513                             &15.57              &0.3315             &0.1298     &0.2230          \\
        PEPLER      &67.54          &0.2920             &0.3183             &0.3052                             &14.18              &0.1476             &\underline{0.1044}     &0.1050  \\
        XRec        &74.53          &0.3946 &0.3506             &0.3730                             &11.45              &\textbf{0.0969}    &0.1048     &\textbf{0.0852}\\
        G-Refer (7B)&\underline{74.91}          &0.3573             &\textbf{0.4264}    &0.3922    &10.88              &0.1050             &\textbf{0.0952}        &\underline{0.0862}     \\
        \rowcolor{gray!20}
        \ours-P     & \textbf{76.25}           &\underline{0.4879}     & \underline{0.3604} 	& \underline{0.4237} 	   & \underline{10.64}           &0.1033 &	0.1153 &	0.0933 	  \\
        \rowcolor{gray!20}
        \ours-L       & 74.32	        &\textbf{0.5005}	&0.3603	&\textbf{0.4298}   &\textbf{10.40}	           &\underline{0.0994}	&0.1160	     &0.0938	 \\

        \midrule
        \multicolumn{9}{c}{\textbf{Google-reviews}}\\
        \midrule
        NRT         &58.27          &0.3509             &0.3495             &0.3496                             &19.16              &0.2176             &0.1267     &0.1571          \\
        Att2Seq     &61.31          &0.3619             &0.3653             &0.3636                             &17.47              &0.1855             &0.1247     &0.1403          \\
        PETER       &65.16          &0.3892             &0.3905             &0.3881                             &17.00              &0.2819             &0.1356     &0.2005          \\
        PEPLER      &61.58          &0.3373             &0.3711             &0.3546                             &17.17              &0.1134             &0.1161     &0.0999          \\
        XRec        &69.12          &0.4546 &0.4069             &0.4311                             &14.24              &\textbf{0.0972} &0.1163     &\underline{0.0938} \\
        G-Refer (7B)&\textbf{71.47}          &0.4253             &\textbf{0.4873}    &\underline{0.4566}        &\textbf{13.46}              &0.1184  &\textbf{0.0872} &\textbf{0.0921} \\   
        \rowcolor{gray!20}
        \ours-P     & \underline{70.35}           &\underline{0.4565} 	&0.4200 	&0.4385 	  & 14.52           &0.1060 	&\underline{0.1130} 	&0.0940  \\
        \rowcolor{gray!20}
        \ours-L       & 69.91           &\textbf{0.4884}    &\underline{0.4259} &\textbf{0.4573}        &\underline{14.23} &\underline{0.1001}     &0.1179    &0.0957           \\
        \bottomrule
    \end{tabular}
    }
\end{table}

\begin{table}[t]
    \centering
    \caption{Ablation study. The best results are highlighted in bold and the worst in \textcolor{red}{RED}. ``RR'' represents the retrieval module, and ``HA'' is the hierarchical aggregation module. ``Random'' indicates that user/item profiles are generated by randomly sampling the user/item reviews.}
    \label{ablation_study}
    \resizebox{0.77\columnwidth}{!}{
    \begin{tabular}{c|c|cc|cc}
        \toprule
        \multicolumn{2}{c|}{Datasets} & \multicolumn{2}{c|}{\textbf{Amazon-books}$\uparrow$}                                                                   & \multicolumn{2}{c}{\textbf{Yelp}$\uparrow$} \\
        \midrule
        \multicolumn{1}{c|}{RR} & \multicolumn{1}{c|}{HA} & $\mathrm{BERT_{score}^{Precision}}$ & $\mathrm{BERT_{score}^{F1}}$  & $\mathrm{BERT_{score}^{Precision}}$ & $\mathrm{BERT_{score}^{F1}}$  \\
        \midrule
        w/o RR & Random   & 0.4570  &	0.4319  &  0.4860    &	0.4192   \\
        w/o RR & w/ HA    & 0.4563  &	0.4347 	&  0.4930    &	0.4250   \\
        w/ RR   & Random  & \textcolor{red}{0.4417}  &	\textcolor{red}{0.4085}   &	 \textcolor{red}{0.4791}    &	\textcolor{red}{0.4191}     \\
        \midrule
        \multicolumn{2}{c|}{\ours}   & \textbf{0.4726} &	\textbf{0.4403}   &   \textbf{0.5031}  &	\textbf{0.4310} \\
        \bottomrule
        
    \end{tabular}
    }
\end{table}

Tab.~\ref{overall-experiments} provides the overall results.
It can be observed that \ours achieves outstanding performance in explanation quality across evaluation metrics based on GPT and BERT. 
XRec and G-Refer take different approaches: the former leverages graph neural networks to capture collaborative filtering information and integrates it into large models, while the latter further exploits graph-structured data using a hybrid retrieval method that combines node-level and path-level strategies to more precisely utilize user-item interaction graphs. 
In contrast, \ours-L, which uses latent representation query, focuses on mining the semantic information in reviews, and achieves notable improvements on the $\mathrm{BERT^{Precision}_{score}}$ metrics across three datasets with increases of 12.6\%, 26.8\% and 7.46\%. 
Meanwhile, comparing with G-Refer on $\mathrm{BERT^{F1}_{score}}$, \ours achieves increases of 2.76\%, 9.59\% and 0.15\%. 
We observe that \ours-P with profile query achieves higher GPT scores compared to \ours-L across all three datasets, with improvements of 2.26\%, 2.60\%, and 0.64\% on Amazon-books, Yelp, and Google-reviews, respectively. However, it performs slightly worse in terms of BERT scores.

\subsection{Ablation Study}
\label{sec:ablation_study}

We provide the ablation study in Tab.~\ref{ablation_study}. 
It is observed that the hierarchical aggregation module improves the performance, which proves that the profile generated by this module exactly extracts the important information from user/item reviews, benefiting the generation step. 
Rarely using review retrieval module without the hierarchical aggregation module get the worst results.
However, when two modules are combined, the best results are obtained. 
This observation indicates that the deviation of profiles introduces noise and makes LLM confused when faced with the contradiction between the profile and the retrieved reviews. 
In contrast, with the combination of the two modules, LLM can better utilize the retrieved reviews according to correct profiles. 
By integrating these two modules, we achieve a compounded benefit that exceeds what each could offer alone.

\begin{figure}
     \centering
     \begin{minipage}{0.49\linewidth}
         \includegraphics[width=1\linewidth]{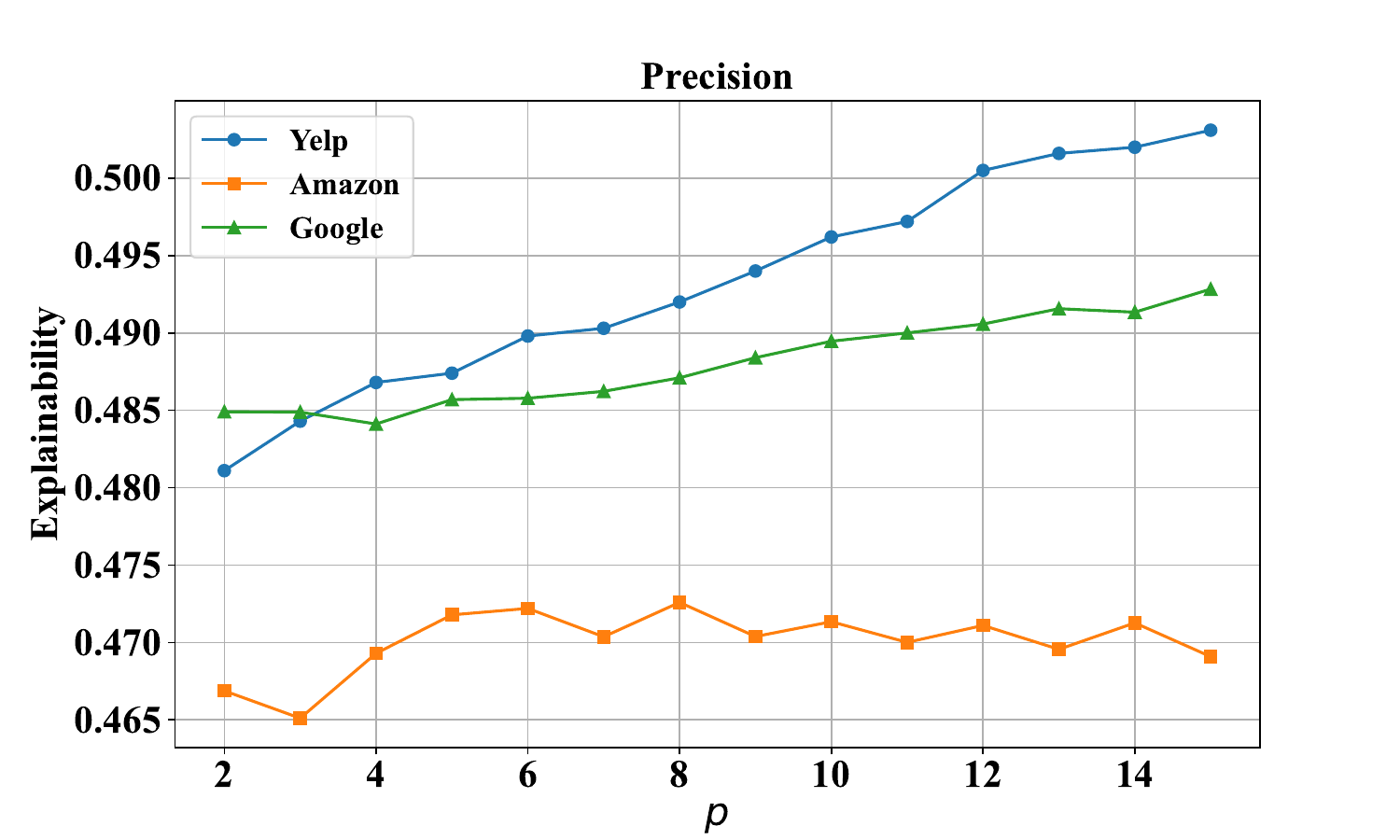}
         \caption{Performance when using different $p$.}
         \label{fig:bert_precision_retrieval}
     \end{minipage}
     \hspace{5pt}
     \begin{minipage}{0.47\linewidth}
         \includegraphics[width=1\linewidth]{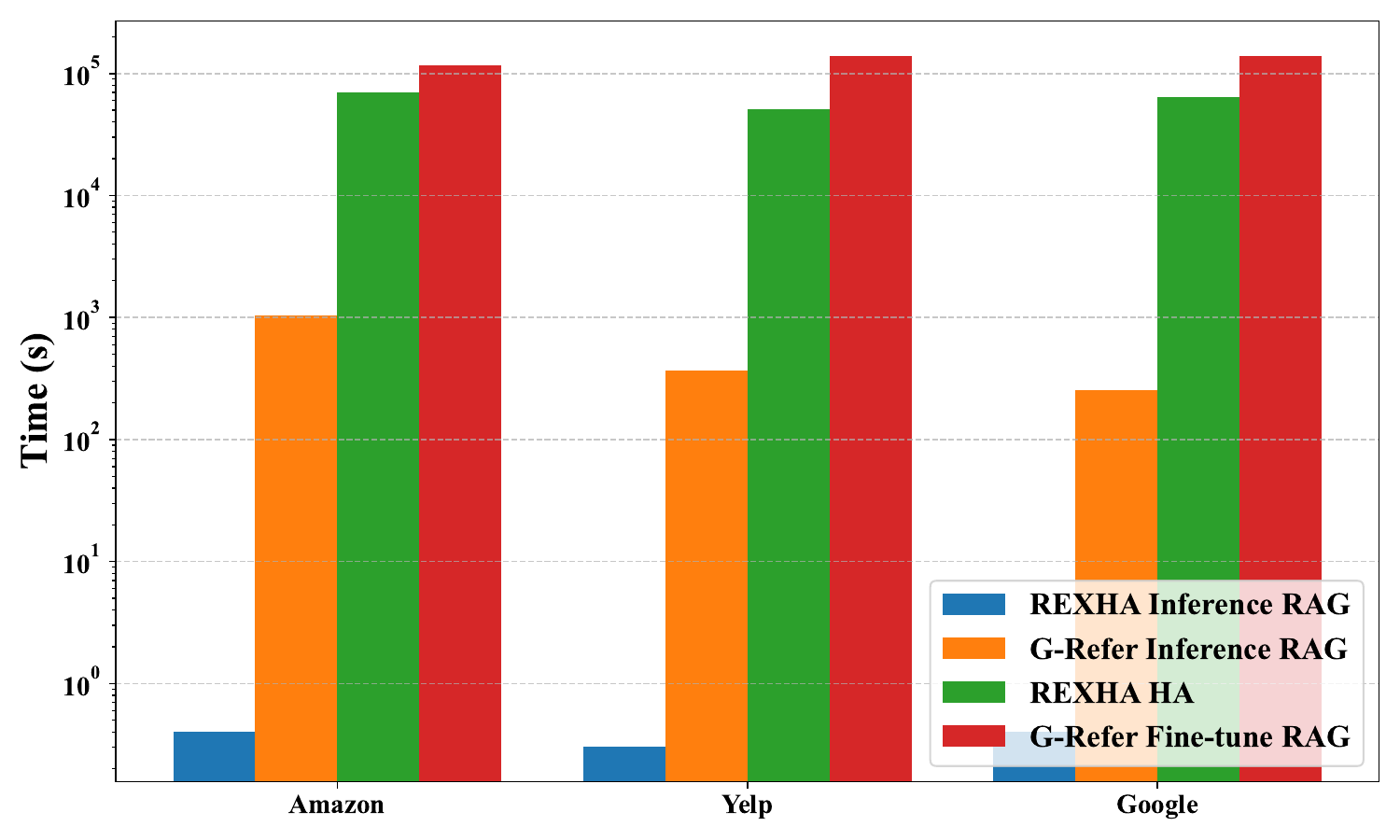}
         \caption{Comparisons of efficiency between \ours and G-Refer.}
         \label{fig:inference_time}
     \end{minipage}
 \end{figure}

\subsection{Analysis of Hyperparameter Sensitivity}

We further investigate how the performance of \ours varies with the number of retrieved reviews $p$. 
Fig.~\ref{fig:bert_precision_retrieval} illustrates the impact of increasing $p$ from 2 to 15 on $\mathrm{BERT_{score}^{precision}}$ and $\mathrm{BERT_{score}^{recall}}$. 
We can observe that on both Yelp, $\mathrm{BERT_{score}^{precision}}$ consistently increases as $p$ grows. When $p>4$, $\mathrm{BERT_{Score}^{Precision}}$ keeps increases on Google-reviews. 
When $p=8$ \ours achieves best performance on Amazon-books.
We notice that $\mathrm{BERT_{score}^{recall}}$ remains relatively insensitive to changes of $p$, indicating that the profile already contains the essential information required for explanation generation.

\subsection{Analysis of Retrieval Efficiency}
\label{sec:retrieval_efficiency}

We evaluate the preprocessing time required by \ours and G-Refer on three datasets. 
Across all three datasets, the retrieval time during the inference stage for \ours is consistently under 1 second, significantly faster than G-Refer, which requires over 4 minutes. This highlights the superior inference efficiency of \ours.
Unlike G-Refer, \ours does not perform any retrieval during the training phase. However, the profile preprocessing step in \ours involves the Hierarchical Aggregation (HA) module, which incurs a relatively high computational cost—taking up to 20 hours. Despite this, it is still considerably more efficient compared to G-Refer, whose total retrieval time during training exceeds 40 hours across all three datasets. These results demonstrate that the HA module, while computationally intensive, remains more efficient than G-Refer's training-time retrieval operations. 

\section{Related Work}

\subsection{Explainable Recommender Systems (ExRec)}

With the rise of LLMs, many recent works leverage their world knowledge to generate more fluent and informative explanations. 
Explainable recommendation (ExRec) enhances user trust by revealing the rationale behind recommendations~\cite{ZhangC20,Ariza-CasabonaB24}, and has received increasing attention. Existing methods fall into three categories: generation-based, extraction-based, and hybrid approaches~\cite{Ariza-CasabonaB24}. Generation-based methods (e.g., NRT~\cite{LiWRBL17}, PETER~\cite{LiZC20}, SEQUER~\cite{Ariza-CasabonaS23}) produce explanations word-by-word from user/item representations. Extraction-based methods (e.g., ESCOFILT~\cite{PugoyK20}, GREENer~\cite{WangCW22}) select sentences directly from reviews. Hybrid methods (e.g., ERRA~\cite{ChengWLZ0LL23}, ExBERT~\cite{ZhanLLLGK23}) integrate both paradigms using retrieval-augmented generation.
With the advent of LLMs, recent works utilize their world knowledge for more fluent and informative explanations. PEPLER applies prompt learning~\cite{LiZC23a}; POD adopts prompt distillation~\cite{LiZC23b}; XRec integrates collaborative signals via a lightweight adaptor~\cite{MaR024}; and G-Refer enhances explanation quality by retrieving structured and semantic CF signals~\cite{LiZLCRKL25}.

\subsection{Retrieval-Augmented Generation (RAG)}

Retrieval-Augmented Generation (RAG) enhances generative models by incorporating external information through retrieval. Recent research focuses on improving retrievers, generators, and their interaction. Recent RAG methods enhance LLMs by retrieving relevant knowledge and incorporating it into the input. A common approach is to concatenate retrieved documents with the original prompt. For example, In-Context RALM~\cite{DBLP:journals/tacl/RamLDMSLS23} appends retrieved content without modifying LLM parameters. SKR~\cite{DBLP:conf/emnlp/WangLSL23} lets the LLM choose between internal knowledge and retrieval. HyDE~\cite{DBLP:conf/acl/GaoMLC23-hyde} generates a hypothetical answer to improve retrieval relevance. RichRAG~\cite{DBLP:conf/coling/0002YWC0D25} retrieves from multiple aspects and fuses the results for richer input.

\section{Limitations}
\label{sec:limitaions}
Although \ours achieves significant improvements in explanation quality and retrieval efficiency, it still has some limitations. For HA module, the time overhead is relatively high, requiring further optimizations such as dynamic pruning of the hierarchical structure based on different user-item pairs to reduce noise. For the retrieval module, we have not yet implemented deeper post-processing of retrieved reviews, including re-ranking the review list and capturing key information.
\section{Conclusion}
\label{sec:conclusion}
We analyze the limitations of existing explainable recommendation methods in terms of profile generation and retrieval efficiency, and propose REXHA to address these challenges. The key components of REXHA include a hierarchical aggregation module that generates comprehensive yet concise user/item profiles, and two different query construction strategies that capture both global and fine-grained features of users and items to retrieve relevant reviews. Extensive experiments demonstrate the efficiency and effectiveness of REXHA, highlighting its promising potential for real-world applications. Future work will focus on deeper utilization of the retrieved reviews and accelerating the profile generation process.

\newpage
\bibliographystyle{unsrtnat}
\bibliography{reference}
\newpage
\appendix

\section{Datasets}
\label{datasets_detail}

\begin{table}[h]
    \centering
    \caption{Statistics of the experimental datasets}
    \label{tab:datasets}
    \renewcommand\arraystretch{1.3}
    \begin{tabular}{cccc}
        \toprule
        Datasets & \#Users & \#Items & \#Interactions \\
        \midrule
        Amazon   & 15,349  & 15,247  & 360,839 \\
        Yelp     & 15,942  & 14,085  & 393,680 \\
        Google   & 22,582  & 16,557  & 411,840 \\
        \bottomrule
    \end{tabular}
\end{table}

We use three public datasets in the experiments and Tab.~\ref{tab:datasets} provides the data statistics. 
\begin{itemize}[leftmargin=*]
    \item \textbf{Amazon-books}
    ~comes from the Amazon book platform and contains user reviews, ratings, metadata, and other information on books. It can reflect user reading preferences and product popularity. 
    \item \textbf{Yelp} comes from the American review website Yelp, where local businesses (such as restaurants and bars) are considered as items. The dataset provides rich information about local businesses, including user review text and ratings under multiple categories.
    \item \textbf{Google-reviews}
    ~comes from user reviews of businesses (such as restaurants, stores, attractions, etc.) on Google, including ratings, review text and timestamps, as well as business metadata.
\end{itemize}

\section{Metrics}
\label{appendix:metrics}
\begin{itemize}[leftmargin=*]
    \item $\mathrm{\mathbf{GPT_{score}}}$~\cite{DBLP:journals/corr/abs-2303-04048} utilizes LLMs to evaluate text quality. We regard GPT-3.5-Turbo ChatGPT as a human evaluator and give task-specific (e.g., summarization) and aspect-specific (e.g., relevance) instruction to prompt ChatGPT to evaluate the generated results of NLG models.
    \item $\mathrm{\mathbf{BART_{score}}}$~\cite{DBLP:conf/iclr/ZhangKWWA20} assesses the similarity between each token in the text by using the contextual embeddings generated by the pre-trained BERT model and calculating the sum of the cosine similarities of the token embeddings between two sentences.

    Given a reference sentence $x=\langle x_1, x_2, \dots,x_n\rangle$ and a candidate sentence $\hat{x}=\langle \hat{x}_1, \hat{x}_2, \dots,\hat{x}_m\rangle$. Bert computes word embeddings sequence for the reference sentence and the candidate sentence as follows:
    \begin{align}
        & \mathbf{x} = \langle \mathbf{x}_1, \mathbf{x}_2, \dots ,\mathbf{x}_n \rangle = \mathrm{BERT}(x=\langle x_1, x_2, \dots,x_n\rangle) \\
        & \hat{\mathbf{x}} = \langle \hat{\mathbf{x}}_1, \hat{\mathbf{x}}_2, \dots ,\hat{\mathbf{x}}_n \rangle = \mathrm{BERT}(x=\langle \hat{x}_1, \hat{x}_2, \dots, \hat{x}_n\rangle)
    \end{align}
    The cosine similarity of a reference token $x_i$ and a candidate token $\hat{x}_j$ is $\frac{\mathbf{x}_{i}^{\top}\mathbf{\hat{x}}_{j}}{\|\mathbf{x}_{i}\|\|\mathbf{\hat{x}}_{j}\|}$. As both embeddings are  pre-normalized, the formula is reduced to the inner product $\mathbf{x}_{i}^{\top}\mathbf{\hat{x}}_{j}$. Then the $\mathrm{BERT_{score}^{Precision}}$, $\mathrm{BERT_{score}^{Recall}}$ and $\mathrm{BERT_{score}^{F1}}$ are computed as follows:
    \begin{align}
          &\mathrm{BERT_{score}^{Precision}}=\frac{1}{|x|}\sum_{x_i\in x}\max_{\hat{x}_j\in\hat{x}}\mathbf{x}_i^\top\mathbf{\hat{x}}_j, \\
          &\mathrm{BERT_{score}^{Recall}}=\frac{1}{|\hat{x}|}\sum_{\hat{x}_j\in\hat{x}}\max_{x_i\in x}\mathbf{x}_i^\top\mathbf{\hat{x}}_j,\\
          &\mathrm{BERT_{score}^{F1}}=2\frac{P_{\mathrm{BERT}}\cdot R_{\mathrm{BERT}}}{P_{\mathrm{BERT}}+R_{\mathrm{BERT}}}.
    \end{align}
\end{itemize}

\section{Baselines}
\label{appendix:baselines}
\begin{itemize}[leftmargin=*]
    \item \textbf{NRT}~\citep{LiWRBL17} tackles both rating prediction and tip generation by learning shared representations from user and item IDs through a joint optimization approach. It utilizes a GRU to produce concise, abstractive tips.
    \item \textbf{Att2Seq}~\citep{ZhouLWDHX17} builds upon an attribute-to-sequence framework where an attention mechanism helps the model focus on relevant input features. A stacked LSTM is employed for decoding the review text.
    \item \textbf{PETER}~\citep{DBLP:conf/acl/LiZC20-peter} introduces a Transformer-based approach that personalizes review generation by linking ID-based representations of users and items with natural language output. Due to dataset limitations, the basic version without auxiliary word features is applied.
    \item \textbf{PEPLER}~\citep{LiZC23a} leverages a pretrained language model to generate textual explanations. It refines this process with techniques like sequential adaptation and regularization to narrow the gap between the prompt structure and the language model's expectations.
    \item \textbf{XRec}~\citep{MaR024} enhances text generation by injecting collaborative filtering signals from GNN-based user and item encoders into every layer of a language model, enabling more contextually relevant and personalized content.
    \item \textbf{G-Refer}~\citep{LiZLCRKL25} strengthens explainable recommendation by retrieving and translating collaborative filtering signals from user-item graphs into human-readable text, enabling large language models to generate personalized explanations through retrieval-augmented fine-tuning.
\end{itemize}

\section{Analysis of Hierarchical Aggregation}
To validate the effectiveness of the Hierarchical Aggregation (HA) method, we compare it with the ``Direct'' method, where the LLM is invoked only once to generate the profile based on all reviews at once. We also compare the performance of Qwen2.5-7B with Qwen2.5-7B-1M, which supports a 1M-token context length.

Tab.~\ref{tab:hierarchical_study} reports the results. 
We observe that Qwen2.5-7B fails to generate profiles using the Direct method since the input exceeds its context length. In contrast, the HA method outperforms the Direct method, achieving improvements of 1.57\%, 0.17\%, and 1.02\% on $\mathrm{BERT_{score}^P}$, $\mathrm{BERT_{score}^{R}}$, and $\mathrm{BERT_{score}^{F1}}$, respectively.
Additionally, we evaluate an alternative approach in which second-layer summaries are directly concatenated to form the profiles. This method achieves slightly higher BERT precision than HA (an increase of 0.08\%), but results in a slight drop in recall (a decrease of 0.19\%).

\begin{table}[h]
    \centering
    \caption{Contrast study for Hierarchical Aggregation module. We evaluate the performance without Review Retrieval module. In this table, ``Directly'' denotes that for all reviews of user/item, LLMs are invoked once to generate profiles, and ``Second Layer'' denotes that sub-nodes of the final node are used as profiles.}
    \label{tab:hierarchical_study}
    \resizebox{0.7\columnwidth}{!}{
    \begin{tabular}{c|c|ccc}
        \toprule
        Models & Type &  $\mathrm{BERT_{score}^P}$ & $\mathrm{BERT_{score}^{R}}$ & $\mathrm{BERT_{score}^{F1}}$  \\
        \midrule
        Qwen2.5-7B-1M & HA        & 0.4712 & 0.3576 &	0.4142  \\
        Qwen2.5-7B-1M & Directly & \textcolor{red}{0.4639} & 0.3570 &	\textcolor{red}{0.4100}  \\
        Qwen2.5-7B-1M & Second Layer & 0.4716 &	\textcolor{red}{0.3557} &	0.4134  \\
        Qwen2.5-7B    & Directly & --- & --- & ---  \\
        Qwen2.5-7B    & HA        & \textbf{0.4930} & \textbf{0.3580} &   \textbf{0.4250}  \\
        \bottomrule
    \end{tabular}
    }
\end{table}

\section{Comparison of Retrieval Results between Latent Representation Query and Profile Query}

\begin{figure}[h]
     \centering
     \begin{minipage}{0.49\linewidth}
         \includegraphics[width=1\linewidth]{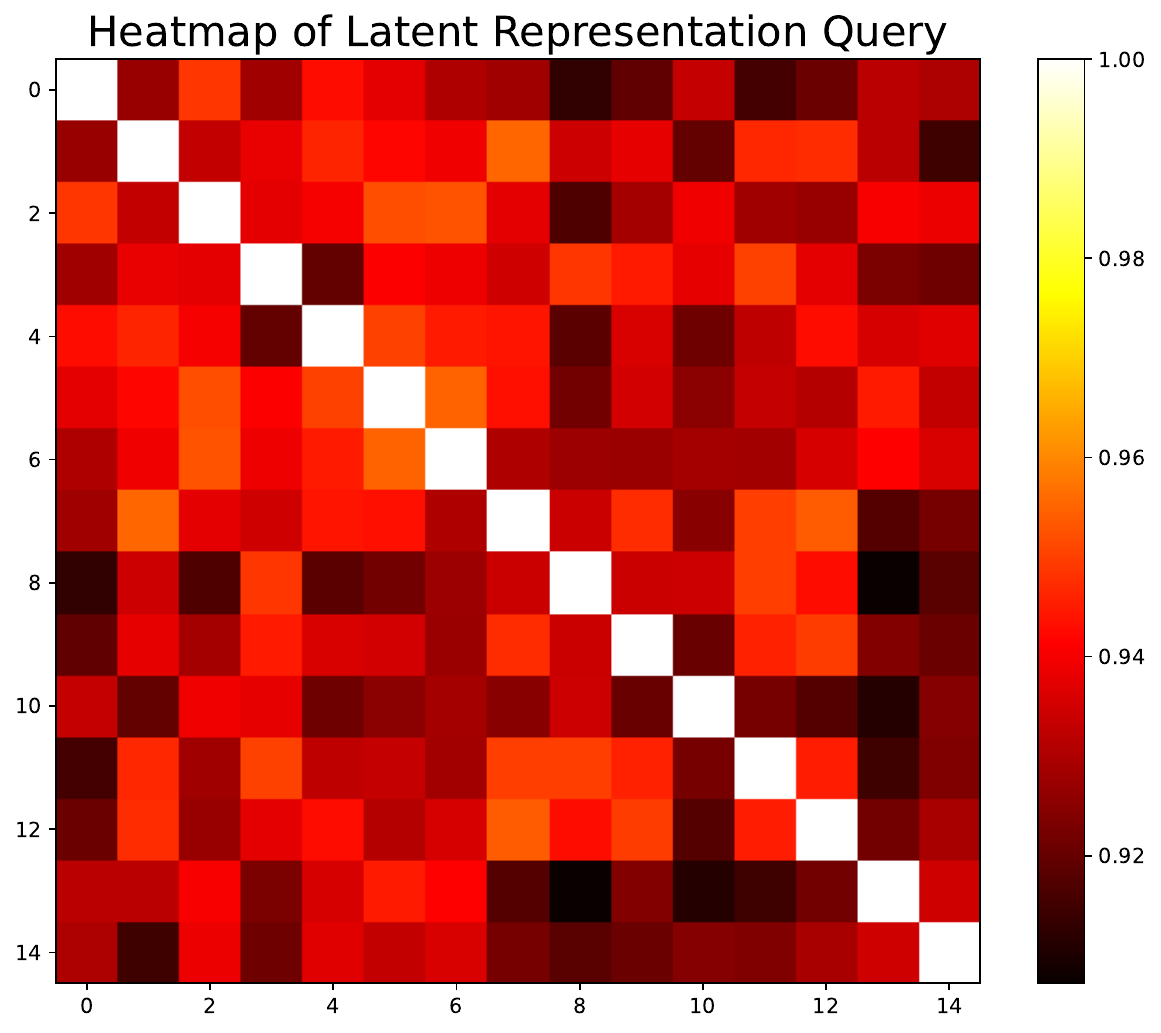}
         \caption{Similarity of each reviews retrieved by latent representation query.}
         \label{fig:heatmap_latent}
     \end{minipage}
     \hspace{5pt}
     \begin{minipage}{0.47\linewidth}
         \includegraphics[width=1\linewidth]{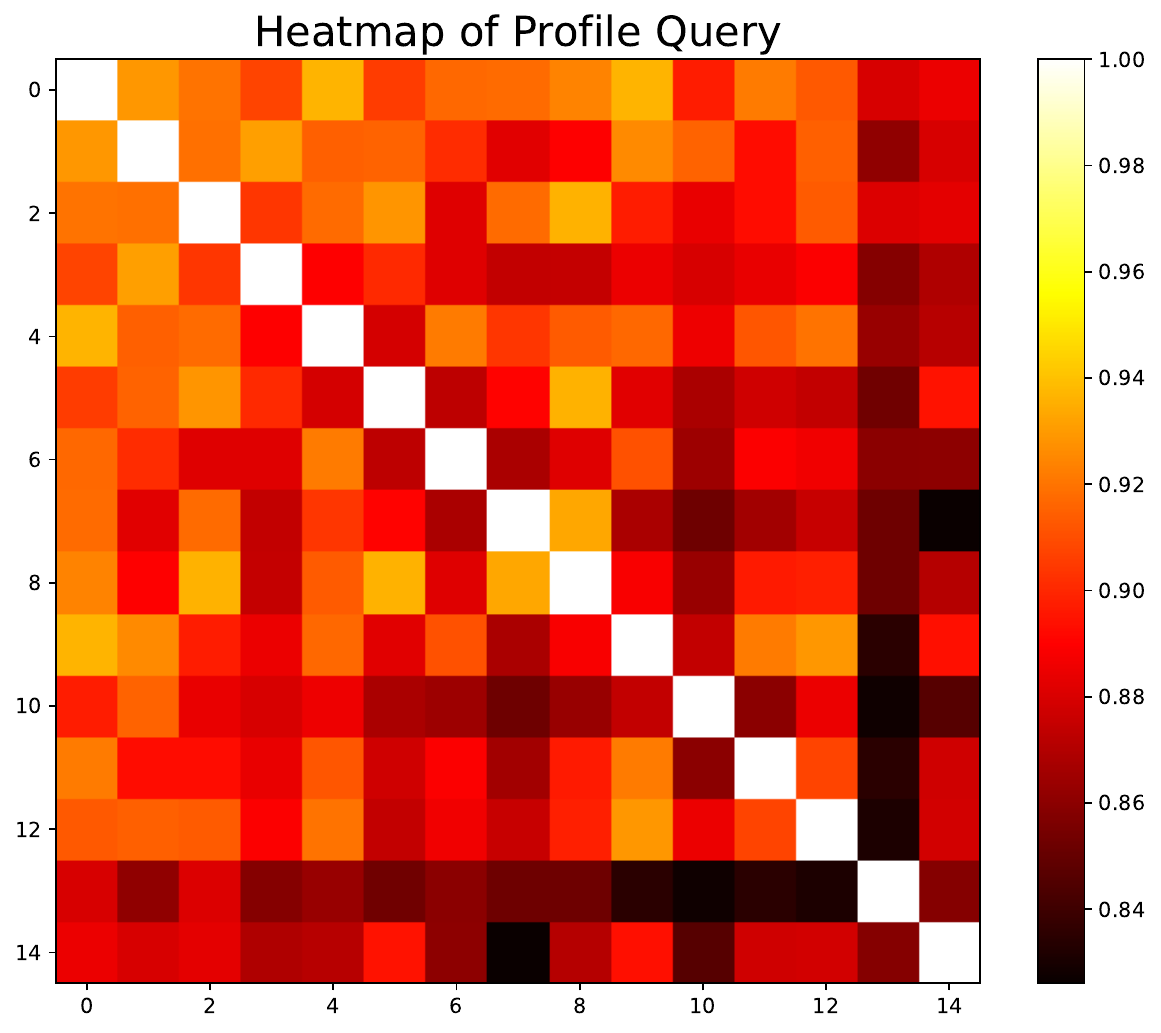}
         \caption{Similarity of each reviews retrieved by profile query.}
         \label{fig:heatmap_profile}
     \end{minipage}
 \end{figure}

To examine the differences between the latent representation query and the profile query, we sampled a set of user-item pairs and calculated the pairwise similarity among the reviews retrieved by each query type. The results are shown in Fig.~\ref{fig:heatmap_latent} and Fig.~\ref{fig:heatmap_profile}. As observed, the reviews retrieved by the latent representation query exhibit lower similarity compared to those retrieved by the profile query, indicating that the former leads to more diverse retrievals. In contrast, the profile query yields more semantically concentrated results, better aligning with long-term user and item preferences.

\section{Prompt Template}
\label{prompt_template}

\begin{figure}[h]
    \centering
    \includegraphics[width=0.8\linewidth]{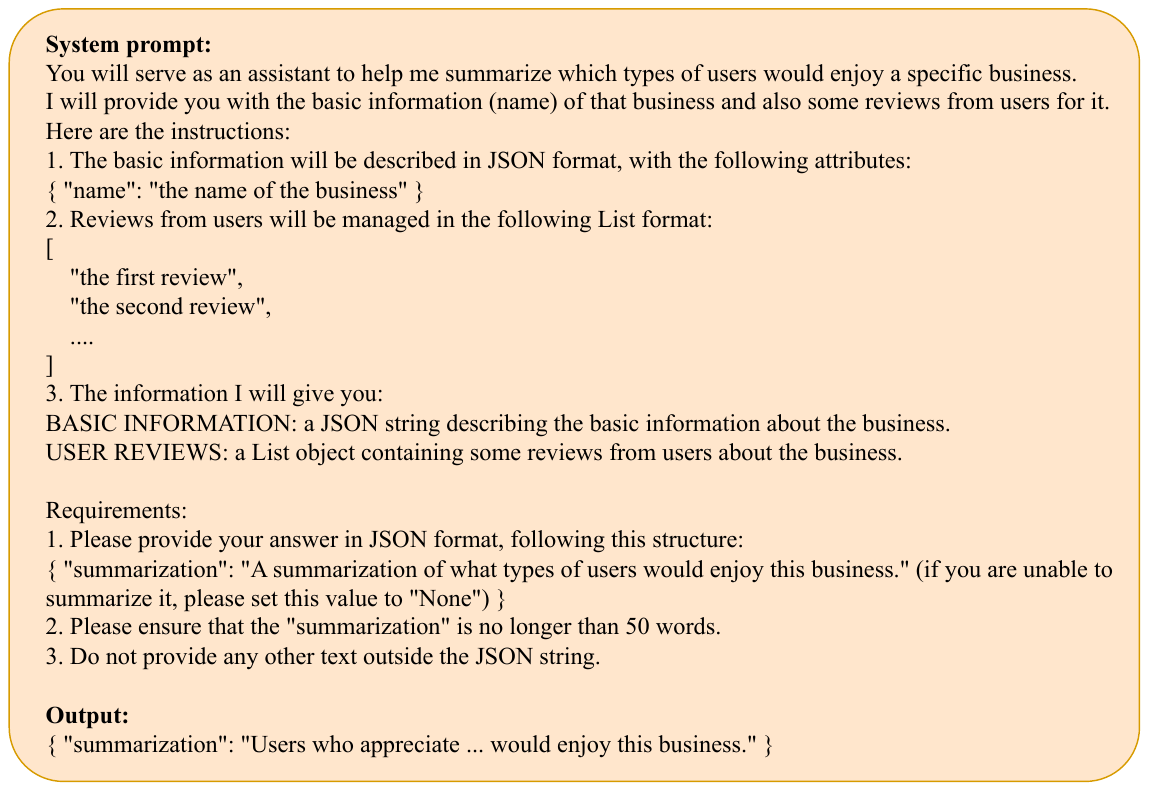}
    \caption{Prompt for Item Profile}
    \label{fig:item_profile_prompt}
\end{figure}

Fig.~\ref{fig:item_profile_prompt} demonstrates how LLMs generate item profiles on Yelp by combining item metadata and user reviews. 
This enables the model to identify key user characteristics associated with the item, offering deeper insights into user preferences and improving the understanding of user-item interactions.

\begin{figure}[h]
    \centering
    \includegraphics[width=0.8\linewidth]{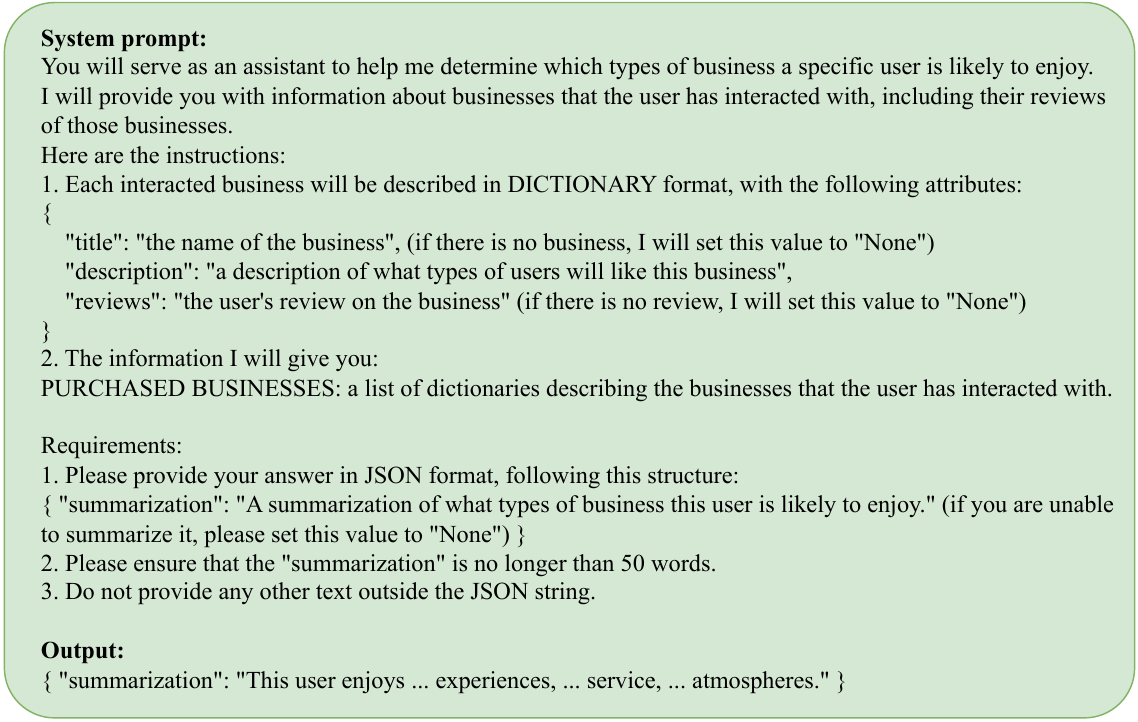}
    \caption{Prompt for User Profile}
    \label{fig:user_profile_prompt}
\end{figure}

Fig.~\ref{fig:user_profile_prompt} presents a method for generating user profiles on the Yelp dataset using LLMs. 
By combining various item metadata, description (i.e., generated item profiles) and user reviews, the system builds detailed user profiles that capture preferences and support personalized recommendations.

\begin{figure}[h]
    \centering
    \includegraphics[width=0.8\linewidth]{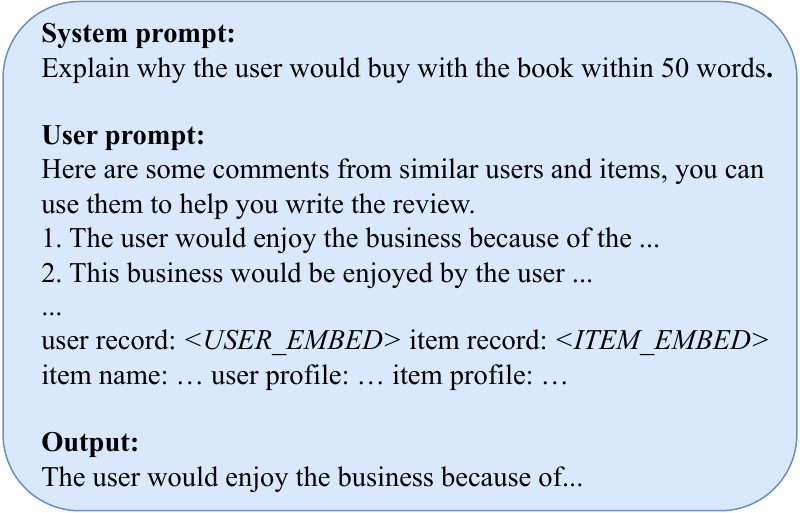}
    \caption{Prompt for Reviews Summarization.}
    \label{fig:prompt_template}
\end{figure}

The generation process employs a structured prompt, as illustrated in Fig.~\ref{fig:prompt_template}, combining multiple data components such as retrieved reviews, user/item embeddings, and user/item metadata. 
The prompt is tokenized and transformed into an embedding space representation. 
To distinguish special tokens within the prompt as distinct entities, we integrate them into the LLM's tokenizer. 
These tokens are subsequently replaced with their corresponding modified embeddings in the final token embedding representation.

\section{Case Study}
We present two cases in Tab.~\ref{case_study_yelp} and Tab.~\ref{case_study_amazon} to demonstrate the effectiveness of our method and illustrate how hierarchical aggregation based profile and retrieved reviews benefit the generated explanations.  
The table only shows some of the reviews on the retrieved content.
We also provide the ground truth explanations and explanations generated by XRec~\citep{MaR024} and G-Refer~\citep{LiZLCRKL25} for comparison.

From Tab.~\ref{case_study_yelp}, we can observe that our method hierarchically aggregates the item profile to highlight ``high quality seafood and steak'' and ``good ambiance''.
This allows our model to integrate the item profile and the retrieved related reviews to generate an explanation that covers both explicit and inferred preferences.
In contrast, XRec incorrectly explains that the restaurant atmosphere is not as good as expected, failing to capture true interests, while G-Refer's explanation is too general, focusing only on common aspects such as "food service" without touching on users' preferences for specific food or atmosphere.
This illustrates that the profile generated by aggregated reviews captures richer and more accurate personalized information than that generated by randomly sampled reviews. 

Tab.~\ref{case_study_amazon} provides another case that leverages retrieved similar reviews to enrich explanations.
While the user profile and item profile simply state their preference for emotionally rich "romance genres", the retrieved reviews reveal deeper elements that appeal to users with similar interests, such as ``intense moments'', ``satisfying endings'', and ``captivating love stories''.
Our model successfully incorporates these signals, capturing both emotional depth and structural highlights that users may value.
In contrast, the explanations generated by XRec are more vague, and G-Refer only captures "intense" information and introduces irrelevant information about the author, which is not relevant to the result.
Together, these two cases demonstrate how retrieved similar reviews and hierarchical aggregate profiles complement each other to produce accurate, diverse, and user-consistent explanations.

\begin{table}[ht]
    \centering
    \caption{A case from Yelp, which shows how \ours leverages HA profile to generate better explanations.}
    \label{case_study_yelp}
    \small
    \resizebox{1\columnwidth}{!}{
    \begin{tabular}{p{0.12\columnwidth}p{0.88\columnwidth}}
        \toprule
        \multicolumn{2}{c}{\textbf{Inputs}} \\ \midrule
        \textbf{User Profile} &  Enjoys craft beers, lively atmospheres, diverse food and drink options, and family-friendly attractions with interactive experiences. \\
        \midrule
        \textbf{Item Profile} & Users who enjoy \textcolor[HTML]{B22222}{high-quality seafood and steaks}, seek good service and value, and appreciate unique dining experiences will enjoy McCormick \& Schmick's. Those looking for affordable yet tasty options and \textcolor[HTML]{B22222}{a good ambiance} during special event days will also find it enjoyable.
        \\
        \midrule  
        \textbf{Retrieved Reviews} & ["The user would enjoy the business because it offers a wide variety of options from cheap produce, meat, and \textcolor[HTML]{B22222}{seafood to diverse dining choices} with the added convenience of a bar, making it a great place for a fun day out with friends.", \par
        "The user would enjoy the business for its diverse beer selection, great customer service, and delicious food options, making it a welcoming and enjoyable spot for sampling different beers and relaxing outdoors.", \par
        "The user would enjoy the business for its interesting beer selection, tasty food like fish and chips, and a cozy atmosphere, despite its modest setting, making it a great spot to explore different dishes and enjoy a casual dining experience.", ... ] \\ \midrule
        \multicolumn{2}{c}{\textbf{Explanations}} 
         \\ \midrule 
        \textbf{\underline{Ground Truth}} & The user would enjoy the business for its \textcolor[HTML]{B22222}{upscale ambiance, delicious steak and seafood options}, excellent cocktails, and the chance to spot a celebrity, making it a versatile spot for meeting friends, dates, or work colleagues.
         \\ \midrule
        \textbf{\underline{XRec}} & The user would enjoy the business for its great food and service, even though the atmosphere is not as good as expected. The menu has a variety of options, and the crab cakes are a highlight. The server was attentive and helpful, making the dining experience enjoyable overall. \\ \midrule
        \textbf{\underline{G-Refer}} & The user would enjoy the business for its delicious food, especially the crab cakes, and the attentive service provided by the server, making the dining experience enjoyable and memorable. \\ \midrule
        \textbf{\underline{\ours (ours)}} & The user would enjoy the business for its \textcolor[HTML]{B22222}{delicious food options, including seafood and steak, and the great atmosphere}, making it a perfect spot for a special occasion or a casual dining experience. \\ 
        \bottomrule
    \end{tabular}
    }
\end{table}

\begin{table}[ht]
    \centering
    \caption{A case from Amazon, which shows how \ours generates better explanation with HA profile and retrieved reviews.}
    \label{case_study_amazon}
    \small
    \resizebox{1\columnwidth}{!}{
    \begin{tabular}{p{0.12\columnwidth}p{0.88\columnwidth}}
        \toprule
        \multicolumn{2}{c}{\textbf{Inputs}} \\ \midrule
        \textbf{User Profile} &  This user prefers emotionally rich, character-driven stories with strong emotional impacts, often in \textcolor[HTML]{B22222}{romance genres}. \\
        \midrule
        \textbf{Item Profile} & Fans of \textcolor[HTML]{B22222}{romantic suspense}, emotional stories, and captivating duets interested in thrilling, suspenseful tales with deep characters and unexpected twists would buy this book.
        \\
        \midrule  
        \textbf{Retrieved Reviews} & ["The user would buy the book because they enjoy a \textcolor[HTML]{B22222}{captivating love story} with complex characters and passionate dynamics, as well as the emotional roller coaster ride provided by the book's plot and the strong emotional investment in the characters.", \par
        "The user would buy the book because it features captivating characters that evoke deep emotional connections, a compelling storyline that elicits \textcolor[HTML]{B22222}{intense reactions}, and \textcolor[HTML]{B22222}{a satisfying ending} that leaves readers feeling ecstatic and emotionally fulfilled.", \par
        "The user would buy the book because it provides a captivating and emotional \textcolor[HTML]{B22222}{love story}, filled with \textcolor[HTML]{B22222}{intense moments} and deep character development, which keeps the reader completely engaged and eager for more.", ... ] \\ \midrule
        \multicolumn{2}{c}{\textbf{Explanations}} 
         \\ \midrule 
        \textbf{\underline{Ground Truth}} & The user would buy the book because it promises a thrilling and unexpected roller-coaster ride of \textcolor[HTML]{B22222}{a love story with intense twists and turns, a satisfying ending}, and relatable themes that make readers question their own love lives.
         \\ \midrule
        \textbf{\underline{XRec}} & The user would buy the book because it offers a compelling and emotionally charged storyline, with well-developed characters and a captivating plot that keeps the reader engaged and invested in the characters' journey, making it a must-read. \\ \midrule
        \textbf{\underline{G-Refer}} & The user would buy the book because they are a fan of the author, Jay Crownover, and have enjoyed her previous works. They are drawn to the \textcolor[HTML]{B22222}{intense} and emotional storyline, the complex characters, and the captivating writing style, making them eager to experience the full impact of the story. \\ \midrule
        \textbf{\underline{\ours (ours)}} & The user would buy the book because it offers a captivating and emotional \textcolor[HTML]{B22222}{love story} with complex characters, \textcolor[HTML]{B22222}{intense moments, and a satisfying ending}, making it a must-read for fans of romance novels. \\ 
        \bottomrule
    \end{tabular}
    }
\end{table}

\end{document}